\def\Version{{7 
    }}






\message{<< Assuming 8.5" x 11" paper >>}    

\magnification=\magstep1	          
\raggedbottom

\parskip=9pt

%

\def\singlespace{\baselineskip=12pt}      
\def\sesquispace{\baselineskip=16pt}      




\font\openface=msbm10 at10pt
 %

\def\Integers      {{\hbox{\openface Z}}}


\font\german=eufm10 at 10pt
\def\Buchstabe#1{{\hbox{\german #1}}}







\def\supp{\mathop {\rm supp }\nolimits}


\def\implies{\Rightarrow}


\def\lto{\mathop
        {\hbox{${\lower3.8pt\hbox{$<$}}\atop{\raise0.2pt\hbox{$\sim$}}$}}}
\def\gto{\mathop
        {\hbox{${\lower3.8pt\hbox{$>$}}\atop{\raise0.2pt\hbox{$\sim$}}$}}}



\def\part{\subseteq}		

\def\braces#1{ \{ #1 \} }


\def\to{\mathop\rightarrow}	

\def\ideq{\equiv}		

\def\less{\backslash}		

\def\alfa{\alpha}


\def\hat{\widehat}		






\font\bmit=cmmib10			
\font\expo=cmmib10 at 10 true pt	

\newfam\boldmath

\textfont8=\bmit 
\scriptfont8=\expo 
\scriptscriptfont8=\expo

  \mathchardef\alpha="710B     
  \mathchardef\beta="710C
  \mathchardef\gamma="710D     
  \mathchardef\delta="710E
  \mathchardef\epsilon="710F   
  \mathchardef\zeta="7110
  \mathchardef\eta="7111       
  \mathchardef\theta="7112
  \mathchardef\iota="7113
  \mathchardef\kappa="7114     
  \mathchardef\lambda="7115
  \mathchardef\mu="7116        
  \mathchardef\nu="7117
  \mathchardef\xi="7118        
  \mathchardef\pi="7119
  \mathchardef\rho="711A       
  \mathchardef\sigma="711B
  \mathchardef\tau="711C       
  \mathchardef\upsilon="711D
  \mathchardef\phi="711E
  \mathchardef\chi="711F
  \mathchardef\psi="7120
  \mathchardef\omega="7121     
  \mathchardef\varepsilon="7122
  \mathchardef\vartheta="7123
  \mathchardef\varpi="7124
  \mathchardef\varrho="7125
  \mathchardef\varsigma="7126
  \mathchardef\varphi="7127

  \mathchardef\imath="717B	
  \mathchardef\jmath="717C	
  \mathchardef\ell="7160



%
 \let\miguu=\footnote
 \def\footnote#1#2{{$\,$\parindent=9pt\baselineskip=13pt%
 \miguu{#1}{#2\vskip -7truept}}}
%
%

\def\linebreak{\hfil\break}

\def\pagebreak{\vfil\break}


\def\BulletItem #1 {\item{$\bullet$}{#1 }}
\def\bulletitem #1 {\BulletItem{#1}}

\def\AbstractBegins
{
 \singlespace                                        
 \bigskip\leftskip=1.5truecm\rightskip=1.5truecm     
 \centerline{\bf Abstract}
 \smallskip
 \noindent	
 } 
\def\AbstractEnds
{
 \bigskip\leftskip=0truecm\rightskip=0truecm       
 }

\def\ReferencesBegin
{
 \singlespace					   
 \vskip 0.5truein
 \centerline           {\bf References}
 \par\nobreak
 \medskip
 \noindent
 \parindent=2pt
 \parskip=6pt			
 }

\def\section #1 {\bigskip\noindent{\headingfont #1 }\par\nobreak\noindent}

\def\subsection #1 {\medskip\noindent{\subheadfont #1 }\par\nobreak\noindent}

\def\reference{\hangindent=1pc\hangafter=1} 

\def\ref{\reference}

\def\journaldata#1#2#3#4{{\it #1\/}\phantom{--}{\bf #2$\,$:} $\!$#3 (#4)}

\def\eprint#1{{\tt #1}}

\def\author#1 {\medskip\centerline{\it #1}\bigskip}

\def\address#1{\centerline{\it #1}\smallskip}

\def\furtheraddress#1{\centerline{\it and}\smallskip\centerline{\it #1}\smallskip}

\def\email#1{\smallskip\centerline{\it address for email: #1}}

\def\PrintVersionNumber{
 \vskip -1 true in \medskip 
 \rightline{version \Version} 
 \vskip 0.3 true in \bigskip \bigskip}

\def\REMARK{\noindent {\csmc Remark \ }}

\font\titlefont=cmb10 scaled\magstep2 

\font\headingfont=cmb10 at 12pt

\font\subheadfont=cmssi10 scaled\magstep1 

\font\csmc=cmcsc10  


\def\EA{\Buchstabe{A}}		
\def\Z2{\Integers_2}
\def\f{\phi}			
\def\CE{{\EA^*}}		
\def\Om{\Omega}		

\def\Abar{{A'}}
\def\Bbar{{B'}}

\def\Star#1{#1^*}
\def\xor{\mathop{\rm xor}\nolimits} 
\def\and{\mathop{\rm and}\nolimits} 
 



\phantom{}


\PrintVersionNumber


\sesquispace

\centerline{{\titlefont An exercise in ``anhomomorphic logic''}\footnote
 {$^{^{\displaystyle\star}}$}%
{To appear in a special volume of {\it Journal of Physics}, edited by
 L. Diosi, H-T Elze, and G. Vitiello, and  devoted to the
 Proceedings of the DICE2006 meeting,  
  held September 2006, in Piombino, Italia.
  \eprint{arxiv quant-ph/0703276}}}  

\bigskip


\singlespace			        

\author{Rafael D. Sorkin}
\address
 {Perimeter Institute, 31 Caroline Street North, Waterloo ON, N2L 2Y5 Canada}
\furtheraddress
 {Department of Physics, Syracuse University, Syracuse, NY 13244-1130, U.S.A.}
\email{sorkin@physics.syr.edu}

\AbstractBegins                              
A classical logic exhibits a threefold inner structure comprising 
an algebra of propositions $\EA$, 
a space of ``truth values'' $V$, and 
a distinguished family of mappings $\f$ 
from propositions to truth values.
Classically $\EA$ is a Boolean algebra, $V=\Z2$, and the admissible maps
$\f:\EA\to\Z2$ are {\it homomorphisms}.  If one admits a larger set of
maps, one obtains an anhomomorphic logic that seems better suited to
quantal reality (and the needs of quantum gravity).  I explain these
ideas and illustrate them with three simple examples.
\AbstractEnds


\sesquispace
\vskip -30pt

\section{}   
From a certain point of view, the phrase ``classical logic'' should be
used in the plural, not the singular, because the things with which
logic deals depend on the ``domain of discourse'', and this can vary
both with time and with the ``system'' one has in mind.  
To each such domain corresponds 
its own {\it Boolean algebra}, namely the
algebra $\EA$ of all ``questions'' one may ask about the
system.\footnote{$^{1}$}
{Instead of ``question'' one also says ``predicate'', ``proposition'' or
 ``event''.  I will use these terms interchangeably, and will sometimes
 refer to $\EA$ as the ``event algebra'', for lack of a better term.}
But a domain of questions 
together with
 rules for combining them via {\it and},
{\it xor}, {\it not}, etc, is not all there is to a logic. 
In addition, one has the space of ``answers''\footnote{$^{2}$}
{also called ``truth values''.}
that a question may have, which classically is 
$\Z2=\braces{0,1}=\braces{false,true}$.
And one has also the space of allowed ``answering maps'' $\f:\EA\to\Z2$, 
which classically coincides with the space of {\it homomorphisms} from the
algebra $\EA$ to the algebra $\Z2$.
To provide  such a $\f$ 
(which I will refer to as a {\it co-event} since it takes events to scalars)
is to answer every conceivable question in $\EA$, 
and consequently to give as full a description of
the corresponding  reality as is classically possible.  

Seen in this way, logic has a threefold character, and one might
think to generalize the classical setup by altering any one of its
basic ingredients: 
the algebra of propositions $\EA$, 
the space of truth-values $\Z2$, or 
the possible maps $\f$.
I have proposed elsewhere [1]
that the paradoxical features of the quantum world can be understood if
one fastens on the last of these three possibilities,
modifying the
nature of the
 co-events 
without
disturbing either
 the space of truth-values
or
 the Boolean character of
the 
event algebra
 $\EA$.
The change wrought by
such a modification
can be thought of in different ways.  
One might say that
reality
remains what 
 it was
for classical logic
(namely an individual ``history'', in the sense of a collection of
particle worldlines, for example\footnote{$^{3}$}
{I am emphatically adopting a ``histories standpoint'', according
 to which reality has a ``spacetime'' nature, not a ``spatial'' one.}), 
but 
one accepts that
it 
will sometimes
manifest
contradictory attributes; 
or one might say 
(less provocatively?)
that the
conception of reality is no longer that of an individual history, 
but rather of a ``non-classical'' co-event $\f:\EA\to\Z2$.

By terming  $\f$ ``non-classical'', 
I merely mean that it is no longer required to be a homomorphism.  
According to classical logic, the truth or falsity of a
compound proposition $P$ like `$A$ or $B$' is unambiguously determined by
the truth or falsity of its individual constituents.  In algebraic
language, if we know $\f(A_i)$ for all the constituents $A_i$
then we also know
$\f(P)$, and the rules for deriving $\f(P)$ from the $\f(A_i)$ say
precisely that $\f$ is a homomorphism of $\EA$ into $\Z2$.  
In the type of generalization proposed in [1], this is no
longer the case.  
One may therefore call such a generalized logic 
``anhomomorphic''.  
(The name ``quantal logic'' would also 
 be suitable were it not
 already in use for a different sort of structural modification
that replaces
 $\EA$
 by the collection of subspaces of a Hilbert space, 
 which is a  lattice but
 not an algebra at all in the strict sense.\footnote{$^{4}$}
{One might ask with what ``quantum logic'' replaces the other two
 logical ingredients, the space of truth-values and the homomorphism
 $\f$.  As far as I know, this question has never been answered, and
 therefore no definite picture of reality has been given.  Instead, one
 has focused on maps from the lattice to the unit interval that
 generalize the classical idea of {\it probability}, rather than that of
  truth.})
By relaxing the demand that a possible coevent be a homomorphism, one can
accommodate contradictions such as that
of the Kochen-Specker gedankenexperiment,
but of course one needs to limit the coevents in some other manner in
order to arrive at a meaningful logical framework.
Otherwise one would be unable to infer the truth or falsity of any
statement about the world from that of any others,
%
%
and
theoretical predictions would become impossible.  

What condition, then, should replace the requirement that co-events be
homomorphisms?  But first, why do we need to replace it at all?  
The reason, as explained in [1], is that we are assuming that
all dynamical truths can be deduced from the {\it preclusion} of certain
events $A\in\EA$, expressed symbolically as $\f(A)=0$.
Among these events are all those for which 
$\mu(A)=0$, where $\mu$ is the ``quantal measure'' 
furnished by the path integral
(see [1]).
For example in 
a diffraction experiment with silver atoms, 
the event of the atom going  to a dark
part of the interference pattern is precluded.
But in combination with the classical laws of inference
this preclusion rule is too powerful.
It ends up denying events that clearly do happen.
In order to accept all the preclusions furnished by quantum theory,
we must therefore give up some of the laws of classical logic.
But in doing so, it is natural
to be guided by the idea that $\f$ should remain as close  to
being a homomorphism as it can. 

The conditions explored briefly in [1] arose in this manner.
They dealt well with several of the simplest paradoxes, and they had the
virtue of reproducing classical logic when the pattern of preclusions
was classical, but it appears that they are still too restrictive
(specifically in connection with 4-slit diffraction [2]).
In this
paper, I will propose 
a modified
 set of rules, motivated by
the same underlying ideas, and then illustrate the resulting logical scheme by
working out 
some
 simple examples.  
In fact I will describe two different
schemes, 
both of which seem so far to be viable.

In proposing them, I am not trying to claim that  either of the 
new
schemes
will prove to be the last word.  Rather, I feel that further 
experimentation with the rules will have to precede any definitive
formulation.  What I hope will prove to be lasting is the insight that 
an ``objective'' interpretation of quantum mechanics can be founded on
the concept of preclusion, 
provided that one generalizes one's conception
of reality by admitting anhomomorphic co-events into one's logic.

Before turning to the specific schemes, let us look at the question in a
somewhat different way.  The problem we are faced with can be seen as a
clash between two ``tendencies of nature''.
On the one hand, events $E$ of very small measure $\mu(E)$ tend 
not to happen.  (This is one way to think about probability as a
``propensity'', in the classical case.)  On the other hand, nature tends
to avoid ``contradictions'' among the events, which we can express
mathematically by saying that nature 
favors
 coevents 
$\f:\EA\to\Z2$ that preserve the logical operations $and$, $not$,
etc.\footnote{$^{5}$} 
{One could also express this by saying that nature tends to
 observe 
 the
 ``laws of inference'' of classical logic.  In the type of scheme I am
 proposing, 
 there will in general be no universal laws of inference, but
 only concrete inferential relationships that depend on which
 system (and ``initial conditions'') one is dealing with.} 
(This could seem an aesthetic requirement, since a map between two
algebras is ``most naturally a homomorphism''.)
Without seeking a deeper understanding of these two tendencies, we can
just accept them and ask what types of possible realities (as co-events)
they point to, given that the two tendencies oppose each other to some
extent.

If the above viewpoint is valid, then
the coevent $\f$ that actually
occurs
 (i.e. the coevent that describes what actually happens)
 results in part from a balancing of  two opposite
tendencies. How this balancing takes place is something we know only in
part.  But to the extent that we can guess the full scheme, we can
determine which coevents are possible and which are not (or are at least
``almost forbidden'').  To do so would be to complete the path-integral
formalism (which defines $\mu$) by producing a predictive dynamical
scheme  free of reference to ``external observers''.
(One might hope to go still farther by formulating the dynamical laws
directly in terms of the coevent itself, without referring to $\mu$ at
all; but no road toward that goal is discernible at present.)

To help us in our ``balancing act'', we have several guides.  The
recovery of classical logic in the
classical limit (and ultimately of probability) is one such guide.  
Another is the need to deal
adequately with ``product systems'', and still others are the need to
furnish a realistic account of what happens in a measurement and to
respect ``relativistic causality''.  However, the best single source of
inspiration may be the ``quantal antinomies'', by which I mean the
paradoxical experiments and thought experiments that illustrate why it
is so difficult to come up with a satisfying interpretation of the
quantal formalism.  These include the EPRB experiment, where
probabilistic correlations play a role, but there are other examples
in which
logic alone leads to the paradox, and the latter offer easier starting
points for exploring a scheme of ``anhomomorphic logic'' such as I am
proposing.  

Before I propose any specific scheme, it would be a good idea to pause
to discuss the mathematical structures of $\EA$, with respect to
which the notion of homomorphism acquires a meaning.  We will also need
to refer to some of the mathematical structures possessed by the space of
coevents, i.e. by the space of functions from $\EA$ to $\Z2$.  
(I will call this space $\CE$ for lack of a better symbol.)
In the background of both the older and newer schemes is a space
$\Om$ of ``histories'' and a {\it quantal measure} $\mu$ or {\it
decoherence functional} $D$ on that space.  For brevity, I will
not describe 
any of this
in further detail, 
referring
the reader to [1] (and references therein) for more explanation.  
I will call $\Om$ the {\it sample space},
and its elements ``formal trajectories'' 
(by analogy with the case where
$\gamma\in\Om$ is a set of particle worldlines).  
And for simplicity I will
always take $\Om$ to be finite.  

On an ``extensional'' view, the event algebra $\EA$ is simply
(given that $\Om$ is finite) the collection of all subsets of
$\Om$.  (So an element  $A\in\EA$ is just a subset of $\Om$;
``intensionally'' it is the corresponding  ``predicate'' or potential
``property of reality''.)  
As such it supports a multitude of operations
that make it, among other things, a poset, a distributive lattice with
complement, a ring, and an algebra over the finite field $\Z2$.  It is a
poset\footnote{$^{6}$}
{partially ordered set}
in an obvious way, with respect to the ordering given by set
inclusion.  It is a lattice with respect to the operations of
intersection and union.  (Indeed a lattice is a special case of a poset.)
And it becomes a ring if one interprets the product `$AB$' as `$A\cap B$'
and the sum `$A+B$' as 
 the symmetric difference `$A\less B \cup B\less A$'.
(Logically, `$AB$' is `$A \and B$', while `$A+B$' is `$A \xor B$' [$\xor$ being
``exclusive or''].)  
The fact that $A+A=0$ makes $\EA$ not only a ring
but an algebra over the finite field $\Z2$, and the fact that $AA=A$
makes it a Boolean algebra.
Notice
in particular that with these definitions, $\EA$ is a vector space over  $\Z2$,
which is something that could not have been deduced merely from the fact
that it is a lattice with respect to ``and'' and ``or''.  
This is also a good place to point out that the
space $\CE$ of co-events $\f$ is also an algebra 
(in fact a Boolean algebra once again), 
simply because it is a function-space, 
and one can
define addition and multiplication pointwise:
$(\f_1 \f_2)(A)=\f_1(A) \; \f_2(A)$ 
and 
$(\f_1 + \f_2)(A)=\f_1(A) + \f_2(A)$.

The variety of ways in which we can conceive of $\EA$ engenders a
corresponding  variety of notions of what it means for 
a function $\f:\EA\to\Z2$
to be a homomorphism.  All these definitions agree when $\f$
actually is a homomorphism, but when it deviates from being so, they can
give rise to different ways to judge by how much it has deviated.  
This is one source of ambiguity in how best to balance ``preclusivity''
against ``homomorphicity''.  

I will call a coevent $\f$ 
{\it preclusive} if it maps every precluded event $A$ to zero (= false), 
and ``homomorphic'' if it preserves the logical operations, 
or equivalently if 
it preserves algebraic sum and product (and $\f(1)\ideq\f(\Om)=1$).
As explained in [1], it cannot
in general do both at once.  In the schemes to be discussed, $\f$ will
be strictly preclusive,\footnote{$^{7}$}
{It seems harder to relax preclusivity in a controlled way than homomorphicity.}
so it will of necessity be anhomomorphic.
The question then becomes in what sense $\f$ can remain ``close'' to a
homomorphism without being literally so.

In the scheme proposed in [1]
(call it the ``linear scheme''), 
the algebraic aspect of $\EA$'s
structure was taken as primary, meaning that preservation of sum
(linearity) and of product (``multiplicativity'') were the criteria for
it to be homomorphic.  Linearity was retained exactly, but
multiplicativity was dropped.  In place of the latter, $\f$ was to be
unital\footnote{$^{8}$} 
{That $\f$ is unital says merely that $\f(1)=1$.  For a homomorphism
 this is automatic except for the trivial $\f$ that vanishes identically:
 $\f=0$.  With events $A\in\EA$ construed as questions, the event
 $1=\Om\in\EA$ asks ``Does anything at all happen?'', and $\f$ is unital
 iff it answers ``Yes'' to this question.
}
and {\it ``minimal''}.  And of course it was to be {\it preclusive}, as we
are assuming always.
The word minimal here refers to the support of  $\f$, which I will
define in a moment.   One says that $\f$ is minimal if there is no
(preclusive, linear) 
$\f$ with smaller support.  For linear $\f$,
$\f$ is homomorphic iff its support is a single element of $\Om$.  
Thus
one may claim that the smaller its support, the more nearly homomorphic
$\f$ is. 
(The minimal such $\f$ also {\it generate} --- i.e. they span --- the
whole vector space of linear preclusive coevents.)

In order to define support, we need some more notation.
For $\gamma\in\Om$ a formal trajectory, let $\gamma^*:\EA\to\Z2$ be the
``containment'' map defined by 
$\gamma^*(A)=1$ if $\gamma\in A$,
$\gamma^*(A)=0$ if $\gamma\notin A$.
(Instead of thinking of $\gamma$ as an element of $\Om$ one can think of
 it as an
 element of $\EA$ by identifying it with the 
 singleton set $\braces{\gamma}$.
 As such it is what is called an {\it atom}
 of $\EA$, meaning a
 minimal element when $\EA$ is regarded as a lattice.  
 Algebraically, $x$ is an atom if for all $y$, $xy$ is either $0$ or $x$
 itself.
 In general it simplifies the notation to identify $\gamma$ with 
 the corresponding atom $\braces{\gamma}$, 
 and I will normally do so in the sequel.) 
It is easy to verify that $\gamma^*$ is a unital homomorphism of $\EA$
onto $\Z2$.  (It thus may be called a ``classical coevent''.)
In particular it is linear.
Moreover the $\gamma^*$ are a basis for the space ${\cal L}(\EA,\Z2)$ of
linear transformations from $\EA$ into $\Z2$.  Thus every 
$\f\in{\cal{L}}(\EA,\Z2)$
decomposes uniquely as a sum of the form
$$
   \f = \sum_{\gamma\in S} \gamma^*  \ ,
$$
where $S$ is some subset of $\Om$ that one may refer to as the {\it
support} of $\f$, $\supp(\f)$.  
More generally, any mapping $\f$ of $\EA$ into $\Z2$ whatsoever can be
expressed as a polynomial in the classical coevents $\gamma^*$ for
$\gamma\in\Om$.\footnote{$^{9}$}
{This leads to a useful graphical notation in which, 
 for $x,y,z\in\Om$,
 $\Star{x}$ is represented as a vertex or 0-simplex,
 $\Star{x}\Star{y}$ is an edge or 1-simplex, 
 $\Star{x}\Star{y}\Star{z}$ is a 2-simplex, 
 etc. 
 For $\f$ represented in this way and $A\subseteq\Om$, $\f(A)=1$  iff $A$
 contains an odd number of such simplices.}
One may thus extend the notion of support by defining $\supp(\f)$ to be
the set of all $\Star{\gamma}$ occurring in the polynomial.

The two schemes we will explore below are related in opposite ways to
the linear scheme.
In both of them reality can be
identified with a single\footnote{$^{10}$}
{Here I'm imagining that $\EA$ includes all possible predicates,
 including ones pertaining to happenings arbitrarily far in the future
 (or past).  If this is not the case, then one will need to consider
 more than just a single event-algebra $\EA$.  There will then be a
 coevent for each $\EA$ together with coherence conditions among 
 these ``partial coevents''.
 Furthermore, a coevent which is ``minimal'' with respect to one
 such $\EA$ can fail to remain so when restricted to a subalgebra.  This
 can lead to a kind of ``premonition'' phenomenon that shows up, for
 example, in the Hardy gedankenexperiment discussed in [1].}
coevent $\f\in\Star{\EA}$.  
As stated already, $\f$ will be preclusive
in both schemes ($\mu(A)=0\implies\f(A)=0$),
but anhomomorphic.
The first of the two schemes
 will be in a sense complementary to the linear scheme.  
Rather than preserving sum it will preserve product (so I'll call it the
``multiplicative scheme'').
The second scheme will 
be a kind of augmented linear scheme, but it will
preserve strictly neither sum nor product.
For reasons that will become clear, I'll call it the ``ideal-based
scheme''.  Since the multiplicative scheme is easier to define, let's
begin with it.

In this scheme, $\f$ preserves the product, $AB$, which in logical terms
means that it preserves conjunction, `$and$'.  
Let us also exclude the trivial multiplicative coevents 
$\f=1$ and $\f=0.$
The
axioms for the multiplicative scheme will then include:

\noindent \qquad $(i) \quad \f \not= 0, \, 1$

\noindent \qquad $(ii) \quad (\forall A,B\in \EA) \ \f(AB)=\f(A) \f(B)$

\noindent
It is relatively easy 
to work out the 
general form 
 of such a $\f$, 
by asking which events are true according to $\f$, 
i.e. for which $A\in\EA$ one has $\f(A)=1$.
Suppose $A\in\EA$, $\f(A)=1$ 
 and $A\subseteq B$.  
Then, since $AB=A$, we have by multiplicativity,
$\f(A)\f(B)=\f(A)=1\implies\f(B)=1$.
Thus $\f$ is monotone: any superset of a true event is also true.  
Now suppose that $A$ and $B$ are both true with respect to $\f$. 
It follows immediately that they have in common some true ``subevent''
(i.e. subset) $C\subseteq A,B$;
for $A\cap B=AB$ is such a subset and $\f(AB)=\f(A)\f(B)=1\times1=1$.
By induction, there must be a smallest true event $F\in\EA$ with the
property that the other true events 
coincide with
its supersets:\footnote{$^{11}$}
{We are just using the fact that $\f^{-1}(1)$  is a {\it filter},
 when $\f\not=0,1$ is multiplicative.}
$$
           \f(A)=1  \iff  F \subseteq A  \eqno(1)
$$
In this scheme, then, a coevent can be construed as a subset $F$ of
$\Om$,
and the rule (1) tells us that, with respect to a given coevent
$\f$, reality ``has the property $A$''
iff all the atoms in $F$ ``share this property''.
Since (1) is a simple generalization of the above definition of
$\Star{\gamma}$, it is natural to write it as: $\f=\Star{F}$.
One can also see that, for any $F\in\EA$,
$$
           \Star{F} = \prod_{\gamma\in F} \Star{\gamma}
$$
Thus a multiplicative coevent is simply a polynomial that reduces to a
monomial (Graphically it is a single simplex),
and the support of $\f$ is $F$ itself:
$$
        F = \supp(\f)  \ .			\eqno(2)
$$

Finally, remember that we want $\f$ to be as nearly homomorphic as it
can be.  Recalling that $\f$ would be a homomorphism 
if its support were
reduced to a single atom, let us postulate that $F=\supp{\f}$ is as
small as possible.  Our final axioms for this scheme are then:

\noindent \qquad $(iii)$ no precluded $A\in\EA$ contains $F$ as a subset

\noindent \qquad $(iv)$ $F$ is minimal subject to $(iii)$

\noindent
(Axiom $(iii)$ requires that $F$ meet the complement of every precluded
subset.)
I will write $\hat{\EA}$ for the set of all coevents that satisfy these
axioms.  The elements of $\hat\EA$ are thus the ``possible realities'',
or ``possible bundles of attributes of reality'', depending on how we
think about a coevent.

Now consider a two-slit diffraction experiment distilled down to a set
of four trajectories joining either of two ``apertures'', $a_1$, $a_2$
to either of two ``detector locations'' $\ell_1$, $\ell_2$.
(We don't include any actual detectors.)
Calling $s$ the source, we have
$$
 \Om = \braces{sa_1\ell_1, sa_1\ell_2, sa_2\ell_1, sa_2\ell_2}
     \ideq \braces{\gamma_1, \gamma_2, \gamma_3, \gamma_4}  \ .
$$
Suppose that $\gamma_1$ and $\gamma_3$ interfere destructively. 
To each $\gamma_i$ corresponds an amplitude $\alfa_i$
(assuming unitary quantum mechanics), and, respecting unitarity, we may
take these amplitudes to be
$$
          \pmatrix{\alfa_1 & \alfa_2 \cr \alfa_3 & \alfa_4}
          = {1\over \sqrt{2}}
          \pmatrix{1 & 1 \cr  -1  & 1}
$$
The only precluded event is then 
$\gamma_1 + \gamma_3$,
which I have written using the algebraic notation introduced earlier.
(Regarded as a subset of $\Om$ this event would be
$\braces{\gamma_1,\gamma_3}$, 
which, strictly speaking, should be written as
 $\braces{\gamma_1} + \braces{\gamma_3}$.  
But the distinction disappears 
when one identifies
$\gamma_i$ with $\braces{\gamma_i}$, as we are doing.)
In order to be preclusive, $\f$ must therefore
satisfy $\f(\gamma_1+\gamma_3)=0$.

 In the multiplicative scheme, $F=\f^{-1}(1)$ must not be contained
within $\gamma_1+\gamma_3$ (axiom $(iii)$).
Hence it must contain either $\gamma_2$ or $\gamma_4$.
For example, $F=\gamma_2+\gamma_3$ will do, as will 
$F=\gamma_4+\gamma_1+\gamma_3$.  Clearly the {\it minimal} sets of this
sort are just $\gamma_2$ and $\gamma_4$ themselves,
corresponding to the coevents 
$\Star{\gamma_2}$ and $\Star{\gamma_4}$.
These two coevents are the members of $\hat\EA$, 
and both are purely
classical.  In each, the particle goes through a single slit and
continues on to the ``bright spot'', avoiding the ``dark spot''.

A more interesting example is three-slit diffraction, as described in
[1].  For simplicity, let's limit ourselves to three
trajectories, $a$, $b$, $c$, all meeting at a common ``detector
location'', and let the corresponding amplitudes be respectively
$1$, $1$, $-1$.
Now both $a+c$ and $b+c$ 
are precluded, 
whence no classical coevent can be preclusive,
since $a+c$ and $b+c$ together cover all of $\Om$.~\footnote{$^{12}$}
{Proving this by computation provides good practice in the algebraic
 notation (even though the
 set-theoretic notation happens to be considerably simpler in this case). 
 We have in general $X\cup Y=X+Y+XY$.  Hence
 $(a+c)\cup(b+c)$ = $(a+c)+(b+c)+(a+c)(b+c)$ = $a+b+2c+ab+ac+cb+c^2$
 = $a+b+0+0+0+0+c$ = $a+b+c=\Om$.}
Classical logic leads to an impasse in this case.
In the multiplicative scheme,
 $F=\f^{-1}(1)=\supp(\f)$
must contain both $a$ and $b$ in order to avoid falling within either 
$a+c$ or $b+c$.
The only possibilities are $F=a+b$ and $F=a+b+c$,
of which only the former is minimal.
Hence $\hat\EA$ contains a single coevent in this example, namely
$\f=\Star{(a+b)}=\Star{a}\Star{b}$;
and for it, 
$\f(a)= \f(b)= \f(c)= \f(a+c)= \f(b+c)=0$, \ 
$\f(a+b)= \f(a+b+c)=1$.
Thus for example, 
the answer to ``Does the particle go through slit $a$?'' is ``No'', 
but
the answer to ``Does it go through {\it some} slit?'' is ``Yes''.
(One could perhaps imagine these questions in terms of idealized
``observations'' that discover no more or less than the question asked
for; 
but of course it would be wrong to identify such ``observations''
with any ordinary physical operations except in very special cases.) 
In this example, the important thing to notice is that there {\it is} a
viable coevent  --- only one in this case, but nevertheless enough to
avoid the untenable conclusion that classical logic would have reached. 

As a final example, consider one that figured heavily in an earlier
interpretation of the path-integral [3].
Here we have two variables $A$ and $B$, each of which can take only the
value $+1$ or $-1$.  If we write $A$ for the event that $A=1$ and
$\Abar$ for the event that $A=-1$,
and similarly for $B$, then our four-element sample space can be written
as
$$
   \Om = AB + A\Bbar + \Abar B + \Abar \Bbar  \ .
$$
Assume that $A$ and $B$ are perfectly correlated, in the 
dynamical
sense that
$\mu(A\Bbar)=\mu(\Abar B)=0$.
From this it follows as well that $\mu(A+B)=\mu( A\Bbar +\Abar B)=0$,
assuming that $\mu$ is strongly positive.\footnote{$^{13}$}
{Strong positivity of the decoherence functional, as defined in
 [4], holds automatically in 
 unitary quantum mechanics.   Via an analog of the Schwarz inequality,
 it implies $\mu(A\cup N)=\mu(A)$ whenever 
 $N$ has measure zero and is disjoint from $A$.}
So our precluded events are
$A\Bbar$, $\Abar B$, and $A+B$.
(Notice that $A+B$ is the event 
 that $A$ and $B$ are {\it anti}-correlated.) 
With our sample space of 4 elements, we have 
$|\Om|=4$, 
$|\EA|=2^{|\Om|}=16$, 
$|\Star{\EA}|=2^{|\EA|}=65536$,
where $|\cdot|$ denotes cardinality.
Of these 65 thousand or so coevents, 
sixteen are multiplicative, 
since there are $2^4=16$ subsets of $\Om$ 
to play the role of $F=\supp(\f)$.
Which of these sixteen coevents 
are preclusive, and which of the preclusive ones are minimal?
Reasoning as before, we see that $F$ must contain either $AB$ or
$\Abar\Bbar$.  But $\f=\Star{(AB)}$ is already preclusive, as is 
$\f=\Star{(\Abar\Bbar)}$.  Hence these are the only minimal preclusive
multiplicative coevents.  Once again, we've reached the classical
solution:
$$
   \hat\EA = \braces{\Star{(AB)},\Star{(\Abar\Bbar)}}  \ .
$$

Indeed this had to happen, because the pattern of preclusions was
classical in the sense that it could have arisen from a classical
probability-measure $\mu$.  Equivalently, every subset of a precluded
set 
was
 also precluded.  One can prove that when this happens the minimal
$\f$ are just the classical coevents $\Star{\gamma}$, for non-precluded 
$\gamma$.  Hence the multiplicative scheme, like the linear one,
reproduces classical logic when quantal interference is absent. 

Although this example of ``$A$-$B$-correlations'' turned out to be
fairly trivial in the end, it is important because it illustrates how
``anhomomorphic inference'' works.  Classically we can conclude from 
``$A+B$ is false'' that exactly one of $AB$, $\Abar\Bbar$ is true.
Consequently, if $A$ is true then $B$ must be true and $\Bbar$ false
(together with $\Abar$).
 None of these deductions is guaranteed a priori in anhomomorphic
logic.  But our analysis of the above example shows that if not only
$A+B$, but also $\Abar B$ and $A\Bbar$ are false 
(which classically would have followed from $A+B$ false),
then classical logic does apply, 
whence the truth of $A$ does imply that of $B$ and conversely.
[proof: if $\f(A)=1$ then $\f$ cannot be $\Star{(\Abar\Bbar)}$ because 
 $\Star{(\Abar\Bbar)}(A)=0$ since $\Abar\Bbar\not\subseteq A$.  Hence
 $\f=\Star{(AB)}$ (the only other possibility in $\hat\EA$), whence
$\f(B)=1$ and $\f(\Abar)=\f(\Bbar)=0$.]  

Finally, let's turn briefly to the ``ideal-based'' scheme that
generalizes the linear scheme of [1].  For lack of space, I
will only sketch this scheme and its application to the above examples.
Its advantage is its greater flexibility and generality, but its
disadvantage is that it is harder to state.  The basic idea is that $\f$
should  be preclusive (as always), and that the members of $\hat\EA$
should be the ``simplest'' that suffice to generate $\Star{\EA}_0$ as an
ideal.\footnote{$^{14}$} 
{An ideal in an algebra is a linear subspace of the algebra which is
 closed under multiplication by arbitrary algebra elements.
 A subset $S$ of an ideal $I$ generates it if $I$ is the smallest ideal
 including $S$.  It is not difficult to verify that the preclusive
 coevents form an ideal.}
Here $\Star{\EA}_0$ denotes the ideal of all preclusive coevents, and
``simple'' is yet another word meaning ``close to homomorphic''.
In the present context it seems best to interpret ``simplicity'' 
not in terms of the support of $\f$, 
but rather in terms of --- say ---
the number of {\it operations} 
needed to build up $\f$ as a polynomial $p$ in the classical coevents.
Thus defined, simplicity is measured by the sum of the degrees of the
monomials whose sum is $p$.  The smaller is this sum, the ``simpler''
is $\f$. 
We then want to take for $\hat\EA$ the generating set of ``maximum
simplicity'', or perhaps just the unital members of this set.
(In detail, there will be different ways to compare the simplicity of
coevents and sets of coevents;
in addition it could happen that the simplest generating set was not
unique.  None of these potential ambiguities surfaces in any of our
three examples.)

Applying the ideal-based scheme
to our three examples, we find the following (limiting ourselves to
the generators that are unital):

\noindent\qquad  {\it 2-slit}:  $\hat\EA$ = same as above.\footnote{$^{15}$}
{There is also a non-classical generator, $\Star{a}+\Star{b}$, but it's not unital.}

\noindent\qquad  {\it 3-slit}:  
  $\hat\EA = \braces{\Star{a}+\Star{b}+\Star{c}, \Star{a}\Star{b}}$

\noindent\qquad  $A$-$B$-{\it correlations}:  $\hat\EA$ = same as above

\noindent 
Thus in these examples, the ideal-based scheme differs from the multiplicative
scheme only in the three-slit example, where it yields
 a second coevent,
namely the unique coevent produced by the linear scheme of
[1].
With respect to this coevent, $a$, $b$ and $c$  all become true, while $a+b$
becomes false. 

One pleasant feature of the ideal-based scheme is that (as one can
prove) if an event $A\in\EA$ is not precluded then there exists
$\f\in\hat{\EA}$   for which $A$ happens ($\f(A)=1$).

\REMARK
Notice that anhomomorphic logic shifts the emphasis from 
individual histories $\gamma\in\Om$
to the
algebra of predicates $\EA$.  
This shift could seem like a retreat from
reality, but it
is perhaps natural from a
``dialectical'' starting point that takes the whole as prior to its
parts.  
To ``unexamined materialism'', reality is simply a single history
$\gamma\in\Om$, what I called earlier a ``formal trajectory''.  Yet such
an element only begins to take shape when we subdivide the whole into
its (spatio-temporally separated) parts.  To the extent that the
subdivision remains to be completed, as it always must, we recognize
only a ``coarse-grained  world'' corresponding to the subalgebra of
$\EA$ generated by the coarse-grained  variables.
Perhaps such musings lend a greater ``dignity'' to the elements of
$\EA$, as being in some sense logically prior to those of $\Om$. 
Anhomomorphic logic is also ``dialectical'' in a second, more obvious
sense: it ``admits contradictions''.


\bigskip
\noindent
It's a pleasure to thank Fay Dowker and Yousef Ghazi-Tabatabai for
providing some key examples that helped give birth to the multiplicative
and ideal-based schemes.
Research at Perimeter Institute for Theoretical Physics is supported in
part by the Government of Canada through NSERC and by the Province of
Ontario through MRI.
This research was partly supported 
by NSF grant PHY-0404646.

\ReferencesBegin                             

\ref [1]  
Rafael D. Sorkin,
``Quantum dynamics without the wave function''
 \journaldata{J. Phys. A: Math. Theor.}{40}{3207-3221}{2007}
 (http://stacks.iop.org/1751-8121/40/3207)
\eprint{quant-ph/0610204} 
http://physics.syr.edu/~sorkin/some.papers/

\ref [2] Yousef Ghazi-Tabatabai (unpublished)

\ref [3] 
  Rafael D.~Sorkin,
``Quantum Measure Theory and its Interpretation'', 
  in
   {\it Quantum Classical Correspondence:  Proceedings of the $4^{\rm th}$ 
    Drexel Symposium on Quantum Nonintegrability},
     held Philadelphia, September 8-11, 1994,
    edited by D.H.~Feng and B-L~Hu, 
    pages 229--251
    (International Press, Cambridge Mass. 1997)
    \eprint{gr-qc/9507057}

\ref [4] 
Xavier Martin, Denjoe O'Connor and Rafael D.~Sorkin,
``The Random Walk in Generalized Quantum Theory''
\journaldata {Phys. Rev.~D} {71} {024029} {2005}
\eprint{gr-qc/0403085}


\end               


(prog1    'now-outlining
  (Outline 
     "\f......"
      "
      "
      "
   ;; "\\\\message"
   "\\\\Abstrac"
   "\\\\section"
   "\\\\subsectio"
   "\\\\appendi"
   "\\\\Referen"
   "\\\\ref....[^|]"
  ;"\\\\ref....."
   "\\\\end